\documentclass[12pt]{iopart}
\usepackage{graphicx}
\usepackage{hyperref}
\usepackage{multirow}
\usepackage{cite}

\renewcommand{\tabcolsep}{0.3cm}

\def\ua{\uparrow}
\def\da{\downarrow}
\def\ra{\rightarrow}
\newcommand{\ket}[1]{\mbox{$|#1\rangle$}}
\newcommand{\bra}[1]{\mbox{$\langle#1|$}}
\newcommand{\braket}[2]{\mbox{$\langle#1|#2\rangle$}}
\newcommand{\set}[1]{\{#1\}}
\newcommand{\outerp}[2]{\ket{#1}\bra{#2}}
\newcommand{\ora}[1]{{\buildrel{#1}\over\longrightarrow\;}}

\begin{document}

\title[]{Qubit state transfer via discrete-time quantum walks}
\author{\.{I}skender Yal\c{c}{\i}nkaya and Zafer Gedik}
\address{Faculty of Engineering and Natural Sciences, Sabanc{\i} University, Tuzla, 34956, \.{I}stanbul, Turkey}
\ead{iyalcinkaya@sabanciuniv.edu}
\vspace{10pt}
\begin{indented}
\item[]January 2015
\end{indented}

\begin{abstract}
We propose a scheme for perfect transfer of an unknown qubit state via the discrete-time quantum walk on a line or a circle. For this purpose, we introduce an additional coin operator which is applied at the end of the walk. This operator does not depend on the state to be transferred. We show that perfect state transfer over an arbitrary distance can be achieved only if the walk is driven by an identity or a flip coin operator. Other biased coin operators and Hadamard coin allow perfect state transfer over finite distances only. Furthermore, we show that quantum walks ending with a perfect state transfer are periodic.
\end{abstract}

\vspace{2pc}


\section{\label{sec:intro}Introduction}
Quantum state transfer from one location to another is a significant problem for quantum information processing systems. A quantum computer, which consists of different processing units, requires the quantum states to be transferred between its parts. Therefore, quantum state transfer will be an important part of quantum computer design. There are various ways of achieving this task depending on the technology at hand \cite{1}. In this article, we consider two related fields of research, quantum state transfer and quantum walks on one-dimensional lattices. 

Quantum communication through a spin chain was first considered by Bose \cite{2} and since then it has been studied in depth \cite{3,4,5,6,7,8,9,10}. This procedure consists of interacting spins on a chain, whose dynamics is governed by Heisenberg, XX or XY Hamiltonians. Perfect state transfer (PST) through a spin chain, in which adjacent spins are coupled by equal strength, can be achieved only over short distances \cite{11,12}.

Quantum walks (QWs) have been introduced as a quantum analogue of classical random walks. The continuous-time QW has been suggested by Farhi and Gutmann \cite{13} as a quantum algorithm to reach the $n$th level of a decision tree faster than the classical random walk. The discrete-time QW has been introduced by Aharonov \etal \cite{14} where the walker has a larger average path length than its classical counterpart. These properties of QWs have allowed development of new quantum algorithms \cite{15}. Many experimental systems for QWs have been implemented. \cite{16,17,18,19,20,21,22,23,24,25,26,27,28,29,30,31,32}. 

The time-evolution of qubit state transfer through a spin chain can be interpreted as a continuous-time QW and PST is possible over a spin chain of any length with pre-engineered couplings \cite{11,12}. Furthermore, this interpretation can be extended to discrete-time QW with a position-dependent coin operator \cite{33}. PST in quantum walks on various graphs has been studied more specifically for the continuous-time model \cite{34}. High fidelity transfer of specific quantum states on variants of cycles has been reported for the discrete-time QW \cite{35} without considering the internal coin state.

In this article, we show the perfect transfer of an unknown qubit state from one site ($A$) to another ($B$) on one-dimensional lattices in discrete-time QW architecture. We treat the coin as our qubit whose state we aim to transfer. The coin is an internal degree of freedom of the walker, e.g. polarization, 2-energy levels, angular momenta or spin, which moves on discrete lattice sites. At the end of the walk, we apply one more coin operator (\textit{recovery operator}) to achieve PST. The recovery operator is independent of the initial coin state and it can be determined before the walk once a coin operator is chosen.  We study the periodicity of each case where PST occurs and show that for all PST cases the quantum walk is also periodic. Moreover, we show that redefinition of the shift operator which amounts a change in the directions in which the walker can move, may lead to PST with appropriate choices of the coin operator.

This article is organized as follows. In section \ref{sec:dtqw}, we present a brief review of discrete-time QW and introduce spatial and local approaches to the definition of directions for the walker. We define the boundary conditions (N-lines and N-cycles) of the walk. In section \ref{sec:wtrans}, we discuss the transfer of the walker between sites $A$ and $B$ without considering the coin state. In section \ref{sec:recpst}, we introduce the recovery operator and give a precise definition of PST for discrete-time QWs. In sections \ref{sec:nline} and \ref{sec:ncycle}, we obtain the cases where PST occurs for N-lines and N-cycles. In conclusion section, we summarize our results.

\section{\label{sec:dtqw}The discrete-time quantum walk}
\begin{figure}[!ht]
\begin{tabular}{ll}
\hspace{1cm}(a) & (b) \\
\centering
\hspace{1.5cm}\includegraphics[scale=0.65]{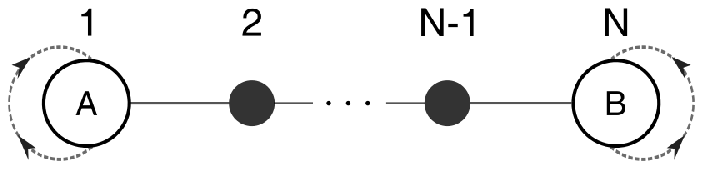} &
\hspace{1.5cm}\includegraphics[scale=0.65]{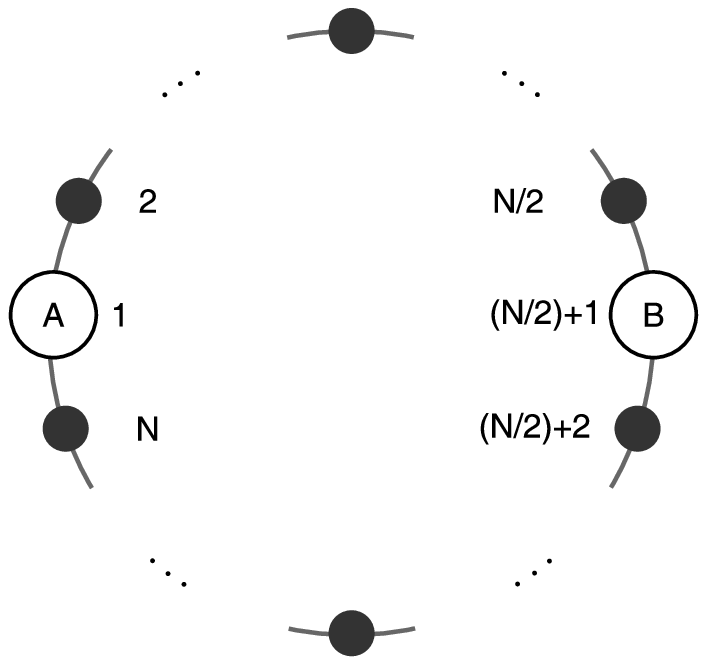} \\
\end{tabular}
\caption{The position space where the quantum walk takes place with (a) reflecting boundary (N-line) and (b) periodic boundary (N-cycle) conditions.  Outermost sites are labelled with $A$ and $B$. Site $A$ is the initial position of the walker and site $B$ is the position where we aim to transfer the coin state.}    
\label{fig:latt}
\end{figure}

One-dimensional discrete-time QW involves two discrete Hilbert spaces. One of them is the position space ${\cal H}_P$ spanned by the basis states $\set{\ket{x} : x \in {\bf Z}}$, and the other one is the coin space ${\cal H}_C$ spanned by basis states $\set{\ket{\ua},\ket{\da}}$. These spaces correspond to the position and the internal states of the \textit{walker}. The total quantum state of the walker is determined by both its coin and position degrees of freedom. In other words, the whole space, ${\cal H}_C \otimes {\cal H}_P $, is spanned by the tensor product of base states which are denoted by $\ket{c,x}$. Time evolution of the walk is governed by a unitary operator which is applied in discrete time intervals (so-called discrete-time) to form the steps of the walk. \textit{One step} is defined by two subsequent unitary operators, \textit{coin operator} which only affects the coin space and \textit{shift operator} which affect both the coin and the position spaces. Thus, one step is given by the unitary operator $\mathbf{U} = \mathbf{S}.(\mathbf{C}\otimes\mathbf{I})$, where $\mathbf{S}$, $\mathbf{C}$ and $\mathbf{I}$ are shift, coin and identity operators, respectively. The most general unitary coin operator can be written as

\begin{equation}
\mathbf{C}=\left( 
\begin{array}{cc}
\sqrt{\rho}		         &	\sqrt{1-\rho}e^{i\theta}     \\
\sqrt{1-\rho}e^{i\phi} & -\sqrt{\rho}e^{i(\theta+\phi)}\\
\end{array} \right),
\label{eq:coinopr}
\end{equation}

\noindent
where $\rho$ gives the bias of the coin, i.e., where $\rho$ and $1-\rho$ are probabilities for moving left and right, respectively. Here, $\theta$ and $\phi$ are the parameters defining the most general unitary operator up to a $U(1)$ phase. The parameters $(\rho,\theta,\phi)$ are chosen arbitrarily at the beginning of the walk and they remain constant during the walk.  For $\rho=1/2$ and $\theta=\phi=0$, we obtain the well-known Hadamard coin operator (unbiased case). The shift operator is given as

\begin{equation}
\mathbf{S}=\ket{\ua} \bra{\ua} \otimes \sum_x \ket{x+1} \bra{x}+\ket{\da} \bra{\da} \otimes \sum_x \ket{x-1} \bra{x},
\label{eq:shift}
\end{equation}

\noindent
where the sum is taken over all discrete positions of the space. The shift operator forces the walker to move in a direction determined by its coin state. The position Hilbert space does not necessarily be infinite and it can be restricted to a finite number of sites $N$ by choosing appropriate conditions for boundaries. In this case, a different definition for the shift operator is required. In figure \ref{fig:latt}, two boundary conditions for the walk are presented and these are the ones that we will use throughout the article. In figure \ref{fig:latt}(a), the lattice with $N$ sites and reflecting boundaries (\textit{N-line}) is represented. Self loops at the boundaries indicate that wave function is reflected after the shift operator is applied, similar to the approach used by Romanelli \etal for the broken links model \cite{36}. The shift operator is of the form

\begin{eqnarray}
\mathbf{S}=&\outerp{\ua}{\da}\otimes\outerp{1}{1} + \outerp{\da}{\ua}\otimes\outerp{N}{N}+ \nonumber \\
&\outerp{\ua}{\ua} \otimes \sum_{x=1}^{N-1} \ket{x+1} \bra{x}+\outerp{\da}{\da} \otimes \sum_{x=2}^{N} \ket{x-1} \bra{x},
\end{eqnarray}

\noindent
where the left (right)-going part at the first (last) site is diverted to the right (left)-going part at the same site to keep the flux conserved. Thus, the shift operator remains unitary. In figure \ref{fig:latt}(b), the lattice with even $N$ sites and periodic boundaries (\textit{N-cycle}) is represented. Here, we simply connect the first and the last sites with one more edge.

\begin{figure}[!ht]
\begin{tabular}{l}
\hspace{2cm}(a) \\ 
\hspace{3cm} \includegraphics[scale=1.2]{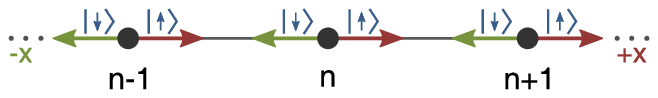} \\
\hspace{2cm}(b) \\ 
\hspace{3.6cm} \includegraphics[scale=1.2]{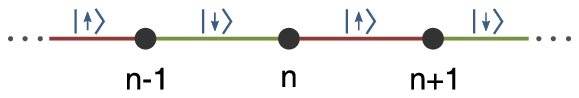}
\end{tabular} 
\caption{Two different approaches which specify the directions in the position space. (a) \textit{Spatial approach}: $\ket{\ua}$ ($\ket{\da}$) state corresponds to $+x$ ($-x$) direction for each site. (b) \textit{Local approach}: Adjacent edges are labelled by different basis states of the coin space in a self-consistent manner. Thus, the walker found at site $n$ takes a step towards $n+1$, and vice versa, if its coin state is $\ket{\ua}$.} 
\label{fig:dirappr}   
\end{figure}

For the walker, directions of motion can be defined in two ways. In the first one, which we shall call as \textit{spatial approach}, the same coin state corresponds to the same spatial direction at every site. Without loss of generality, one can choose the up (down) coin state  to correspond the right (left) spatial direction or clockwise (anti-clockwise) rotation. In the second approach, which we shall call as \textit{local approach}, we assign two orthogonal coin states to the two edges of every site in a self-consistent manner. The discussion here could equivalently be done by redefining the shift operator as well. These approaches are summarized in figure \ref{fig:dirappr}.

A walk can start with any initial state. We will use only localized initial states of the form $\ket{\Psi_0}=\ket{\psi_{0,x}}\otimes\ket{x}$, where $\ket{\psi_{0,x}}=\alpha_{0,x} \ket{\ua}+\beta_{0,x}\ket{\da}$ is the arbitrary initial coin state. The first and the second indices denote the number of steps and site, respectively. After $t$ steps, the initially localized state disperses in the position space and the quantum state of the walker becomes

\begin{equation}
\ket{\Psi_0} \ora{\mathbf{U}^t} \ket{\Psi_t}=\sum_{x}{(\alpha_{t,x}\ket{\ua}+\beta_{t,x}\ket{\da})\otimes\ket{x}}.
\label{eq:evol}
\end{equation}

\noindent
At the end of the walk, the probability of finding the walker at site $x$ is given by summing the probabilities over the coin states

\begin{equation}
P_{t,x}=|\alpha_{t,x}|^2+|\beta_{t,x}|^2.
\label{eq:prob}
\end{equation}

Time evolution which is given in (\ref{eq:evol}) can be written as an iterative map which gives the coefficients of $\ket{\ua}$ and $\ket{\da}$ at any time for any position. First, consider the effect of the coin operator on the wave function projected on a given position state $\ket{x}$ :

\begin{eqnarray}
\ket{\Psi_t'} \ \ora{\mathbf{C}\otimes\mathbf{I}} \ [&(\alpha_{t,x}\sqrt{\rho}\ +\beta_{t,x}\sqrt{1-\rho}e^{i\theta})\ket{\ua}\ +\nonumber \\
&(\alpha_{t,x}\sqrt{1-\rho}e^{i\phi}-\beta_{t,x}\sqrt{\rho}e^{i\theta}e^{i\phi})\ket{\da}]\otimes\ket{x}.
\label{eq:coineffect}
\end{eqnarray}

\noindent
Coefficients of $\ket{\ua}$ and $\ket{\da}$ are the right-going and the left-going probability amplitudes, respectively. Thus the time evolution in (\ref{eq:evol}) can be written as the map

\begin{equation}
\eqalign{
\alpha_{t+1,x}=\alpha_{t,x-1}\sqrt{\rho} + \beta_{t,x-1}\sqrt{1-\rho}e^{i\theta}, \cr
\beta_{t+1,x}=\alpha_{t,x+1}\sqrt{1-\rho}e^{i\phi}-\beta_{t,x+1}\sqrt{\rho}e^{i\theta}e^{i\phi},
}
\label{eq:mapping} 
\end{equation}

\noindent
which is very useful to keep track of the probability flux on the lattice.

\section{\label{sec:wtrans}Transfer of the walker after limited number of steps}
For classical random walks, \textit{recurrence} is defined as the return of the walker to the origin and characterized by P\'{o}lya number \cite{37}. In QW case, the same definition is used and full-revival of the initial quantum state is not necessary \cite{38,39}. Here, we consider the transfer of the walker from site $A$ to site $B$ after limited number of steps without considering its coin state. In other words, we expect that the condition $P_{t,B}=1$ is fulfilled after $t$ steps where $t$ is an integer comparable with the number of lattice sites.

\begin{figure}[!htb]
\begin{tabular}{lll}
(a) & (c) & (e) \\
\includegraphics[scale=0.65]{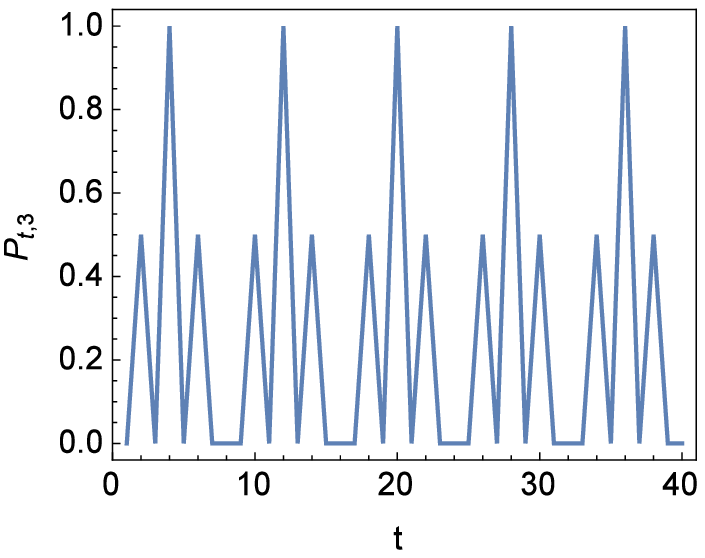} &
\includegraphics[scale=0.65]{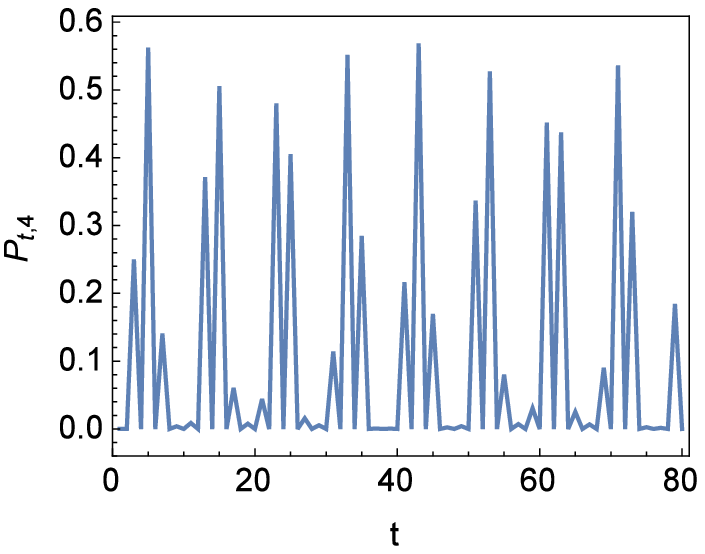} &
\includegraphics[scale=0.65]{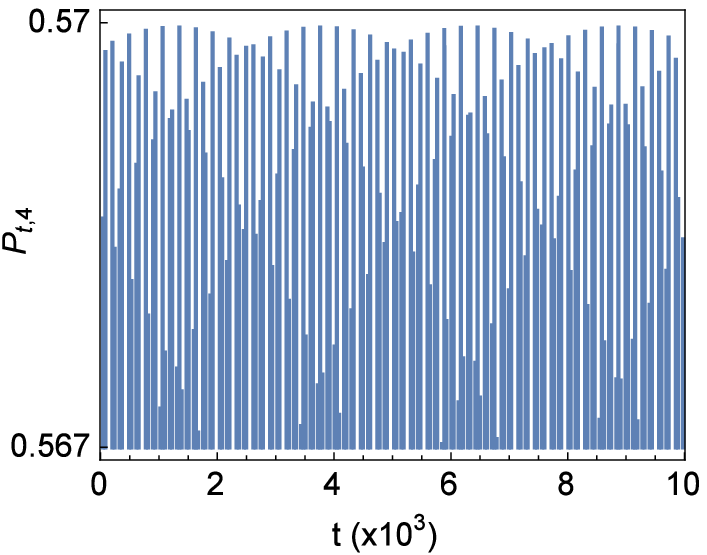} \\
(b) & (d) & (f) \\
\includegraphics[scale=0.65]{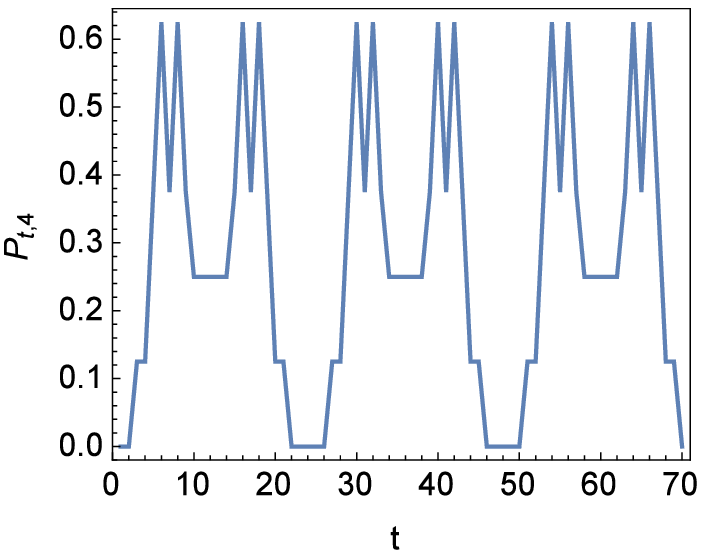} &
\includegraphics[scale=0.65]{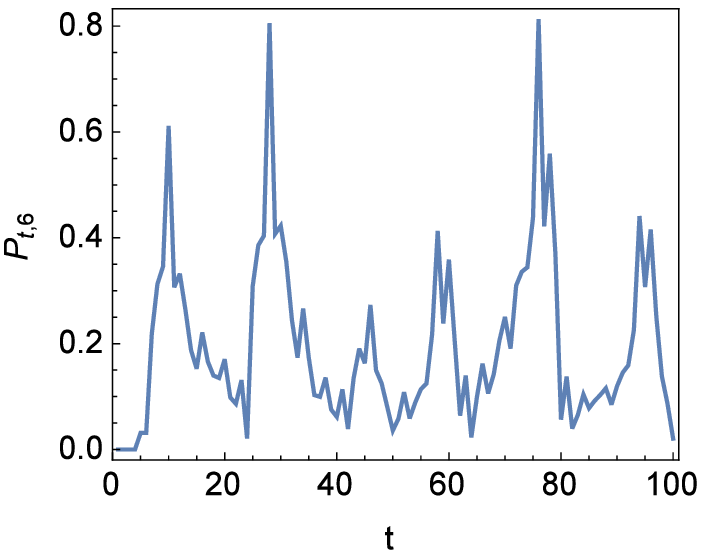} &
\includegraphics[scale=0.65]{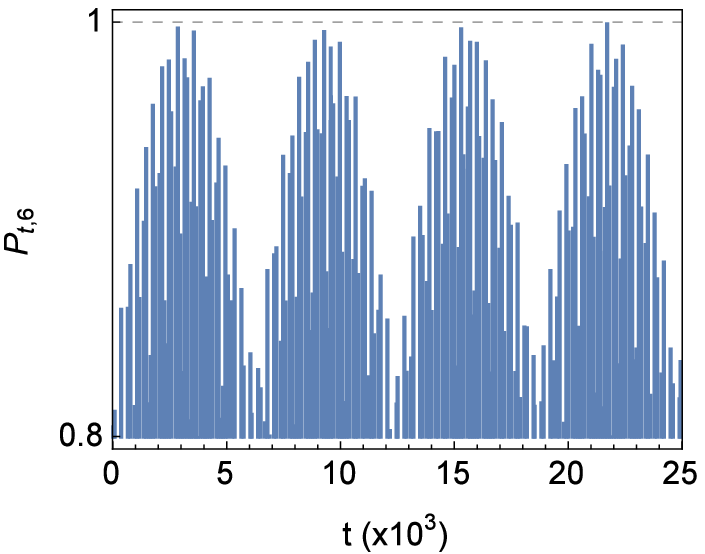}
\end{tabular}
\caption{The probability of finding the walker at the antipodal site $B$ on (a) 4-cycle, (b) 4-line, (c) 6-cycle and (d) 6-line for the Hadamard walk, $\rho=1/2$. The initial coin state is chosen as $\ket{\psi_{0,A}}=(1/\sqrt{2})[\ket{\ua}+i\ket{\da}]$ in each case. Long-time behaviours for (e) 6-cycle and (f) 6-line reveal the recursive behaviour of the probability at site $B$. The lower parts of (e) and (f) are clipped.}    
\label{fig:transfer}
\end{figure}

In figure \ref{fig:transfer}(a), for 4-cycle, we see that the walker appears recursively at site $3$  with a period of $8$ steps for 4-cycle. In figure \ref{fig:transfer}(b), we also observe a recursive behaviour for 4-line but probability never reaches to $1$. It means that it is impossible to transfer the walker from site $1$ to $4$ because $P_{t,4}$ repeats itself in every $22$ steps with $0.625$ maxima. For 6-cycle (figure \ref{fig:transfer}(c)) and 6-line (figure \ref{fig:transfer}(d)), neither the walker can be transferred within the given time intervals, nor a recursive behaviour is observed. However, further analysis for long-time behaviour reveals repetitive patterns which display irregular periodicity \cite{40}. In figure \ref{fig:transfer}(e), we observe that peak values oscillate with quasi-periods of $2412$ and $2698$ steps. The maximum value is $0.57$ and therefore transfer of the walker is impossible. In contrast, figure \ref{fig:transfer}(f) demonstrates that it is quite probable ($\approx 0.99$) to find the walker at site $6$ with quasi-periods of $6416$ and $6016$ steps. In this manuscript, we omit this kind of approximate transfers which appear after very large number of steps because of their unpredictability and experimental inconvenience. Instead, we consider the cases where exact transfer of the walker (e.g., figure \ref{fig:transfer}(b)) is possible in a specific limited time. In figure  \ref{fig:transfer}, we consider the Hadamard walk only for initial coin state $\ket{\psi_{0,A}}=1/\sqrt{2}[\ket{\ua}+i\ket{\da}]$. For PST, we expect to obtain the same behaviour for all initial coin states. Thus, we define our first criterion which we use in our numerical work for PST as follows: \textit{For a given coin operator, the walker has to be transferred to the antipodal site of the lattice after a specific number of steps for all initial coin states.} We examine coin operators whether they satisfy this criterion or not. Once this criterion is satisfied then we examine the final coin state for its similarity to the initial coin state. A precise definition for PST is given in the following section.

\section{\label{sec:recpst}Recovery operator and PST}
The walker is initially localized at site $A$. We are interested in the perfect transfer of walker's coin state from site $A$ to site $B$.  For this purpose we define the \textit{fidelity} at time $t$ and site $B$ by

\begin{equation}
f_{t,B}=|\braket{\Psi_B}{\Psi_t}|,
\label{eq:fidelity}
\end{equation}

\noindent
where $\ket{\Psi_{B}}=\ket{\psi_{0,A}}\otimes\ket{B}$ and $\ket{\Psi_{t}}$ is the quantum state of the walker at step number $t$. $\ket{\psi_{0,A}}$ is the coin state at $t=0$ and site $A$. PST occurs if $f_{t,B}=1$. Thus, we are looking for a class of time evolutions of the form \cite{43} 

\begin{equation}
\ket{\psi_{0,A}}\otimes\ket{A} \ra \ket{\psi_{0,A}}\otimes\ket{B}.
\label{eq:pst}
\end{equation}

\noindent
Since we assume that our initial coin state is unknown, for a given coin operator, the required number of steps to transfer  the coin state have to be the same for all $\ket{\psi_{0,A}}$'s (see the criterion in section \ref{sec:wtrans}). The condition, $A=B$ and $f_{t,B}=1$, implies that the walk is \textit{periodic}, which means that the initial quantum state is completely recovered after $t$ steps up to a phase constant. Periodicity has been first discussed by Travaglione and Milburn \cite{16} for 4-cycle. They have shown the full-revival of the initial quantum state $\ket{\Psi_0}=\ket{\ua}\otimes\ket{0}$ after $8$ steps with Hadamard coin. Later, Treggenna \etal have extended this result by showing that, except 7-cycle, all N-cycles with $N<11$ manifest periodicity with appropriate choices of $(\rho,\theta,\phi)$ for every initial coin state \cite{41}. Dukes has analysed the periodicity of  N-cycles in detail and presented the general conditions for periodicity \cite{42}.  Here, although our main aim is to achieve PST, we will also study the periodicity for each case under consideration. We consider a QW to be periodic, if it is periodic for all initial coin states as in \cite{41}.

As we have discussed in section \ref{sec:wtrans}, for a given number of steps, the walker may appear at $B$ with probability $1$. However, its coin state ($\ket{\psi_{t,B}}$) can be different from the initial one ($\ket{\psi_{0,A}}$). Since  (\ref{eq:coinopr}) includes all possible rotations for a two-dimensional coin, one can transform $\ket{\psi_{t,B}}$ to $\ket{\psi_{0,A}}$  with suitable parameters, ($\rho'$,$\theta'$,$\phi'$). Let us denote this coin operator with primed parameters as $\mathbf{C}_R=\mathbf{C}'\otimes\mathbf{I}$ (\textit{recovery operator}) to distinguish it from the one which governs the walk. Since the initial coin state is unknown, recovery operator have to be independent of the initial coin state. We define this condition as our second criterion for PST as follows: \textit{For a given coin operator and lattice, there should be only one recovery operator which transforms the final coin state to the initial coin state.} In section \ref{sec:nline} and \ref{sec:ncycle}, we show that all cases which satisfy the first criterion (in section \ref{sec:wtrans}) also satisfy the second one. Thus, once we decide on the coin operator which we will use for the walk, we can also determine the recovery operator which will be applied at the end of the walk to achieve PST. This PST scheme can be summarized as

\[
\ket{\psi_{0,A}}\otimes\ket{A}\ \ora{\mathbf{U}^t} \ \ket{\psi_{t,B}}\otimes\ket{B} \  \ora{\mathbf{C}_R} \ \ket{\psi_{0,A}}\otimes\ket{B}.
\] 

In our calculations, $A$ and $B$ are chosen as the outermost sites on the lattice. These are $1$st and $N$th sites for the N-line, $1$st and $(\frac{N}{2}+1)$th sites for the N-cycle with even $N$, respectively. We do not analyse N-cycles with odd $N$ for PST, since we cannot assign a unique $A$ and $B$ pair. First we have numerically determined all cases where the walker is found with probability $1$ at $B$ for all Bloch states (the first criterion). We have restricted the lattice size to $N<11$ if $\rho\neq1$. Without loss of generality, we have also restricted the coin operator to $\phi=0$ \cite{41}. Then, we have analytically studied these cases for their aptness to periodicity and PST  (the second criterion), by using (\ref{eq:mapping}). 

\begin{figure}[!htb]
\begin{tabular}{ll}
\hspace{1cm}(a) & (b) \\
\centering
\hspace{1.5cm}\includegraphics[scale=0.4]{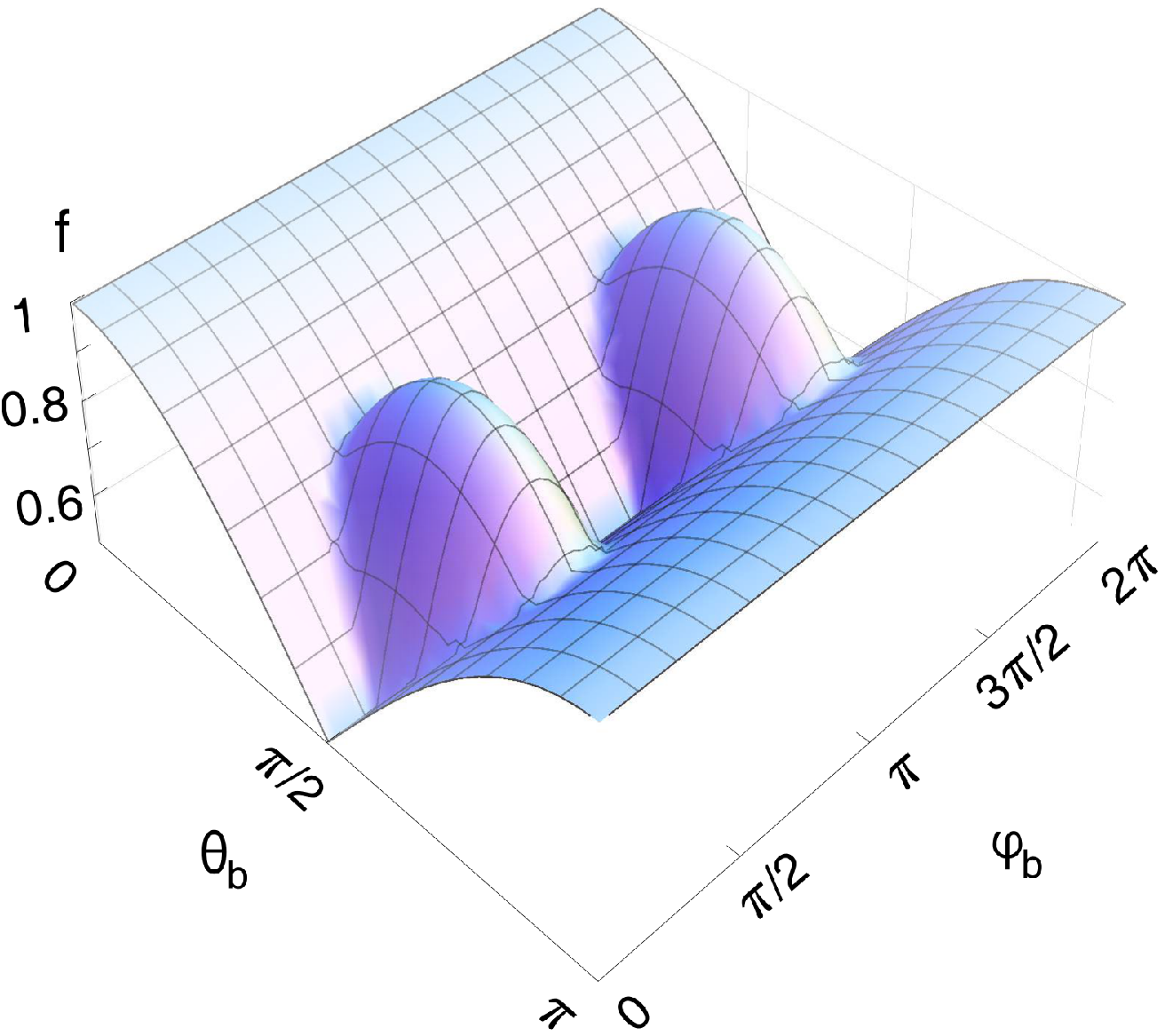} &
\hspace{1.5cm}\includegraphics[scale=0.4]{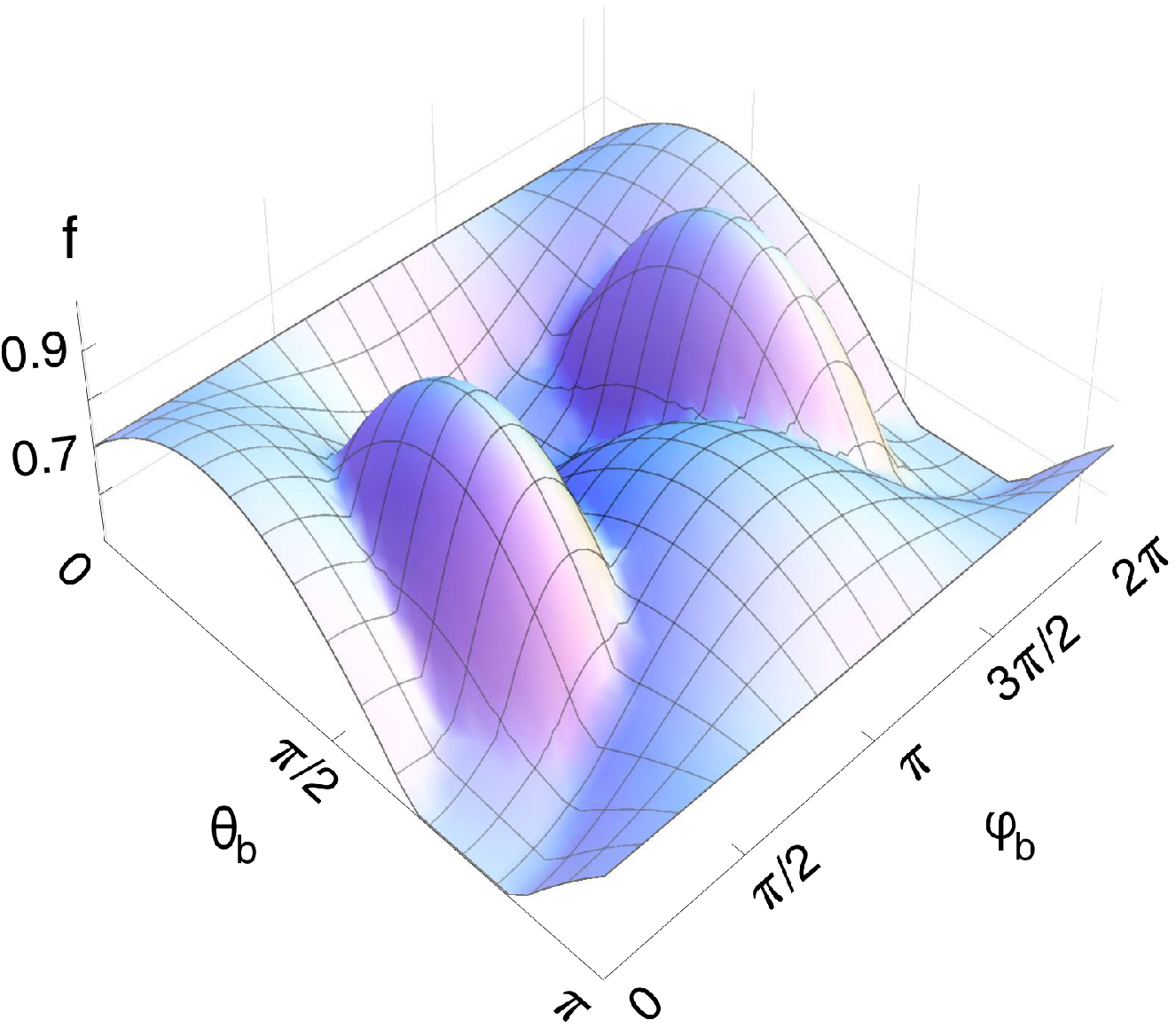} \\
\end{tabular}
\caption{Fidelities of the initial coin states on 2-line. $1$st and $2$nd sites are chosen as $A$ and $B$, respectively. The $(\theta_b,\phi_b)$ plane represents the initial coin states on Bloch sphere and $f$ is the maximum fidelity over time. (a) Identity coin operator is used. Independent of $\phi_b$, the coin states $\theta_b=0$ and $\theta_b=\pi$ are transferred perfectly. (b) Hadamard coin operator is used and no PST is observed within a limited time interval.}
\label{fig:fidelities}
\end{figure}

For PST, without any knowledge about the initial coin state, one should be able to transfer all coin states with $f_{t,B}=1$. However, an arbitrary coin operator and an arbitrary lattice do not provide QWs which allow PST in general. In figure \ref{fig:fidelities}, two specific examples are demonstrated. These are the numerical analyses of fidelity distributions over initial coin states for 2-line. In figure \ref{fig:fidelities}(a), it can be seen that only limited number of initial coin states are transferred perfectly for identity coin operator. In figure \ref{fig:fidelities}(b), Hadamard coin is used and there is no PST at all. Further analysis for the other coin operators and lattices give similar results except the 4-cycle which will be explained in section \ref{sec:ncycle}. 

\section{\label{sec:nline}PST on N-lines}
\subsection{\label{sec:nline1}Case: \texorpdfstring{$\rho\neq1$}{rho!=1}}
In table \ref{tb:2line}, the cases where the walker is found with probability $1$ on the 2-line are given. The cases with coin state $\ket{\psi_{0,1}}$ manifest periodicity. To achieve a PST, we consider the other cases where the total state is

\begin{equation}
\ket{\Psi_t}=[-\beta\ket{\ua}+\alpha\ket{\da}]\otimes\ket{2}.
\label{eq:wf2line}
\end{equation}

\begin{table}[h!]
\caption{\label{tb:2line}For the 2-line, these are the cases where the walker is found with probability 1. The other parameters of the coin operator are chosen as $\theta=\phi=0$.}
\begin{indented}
\item[]\begin{tabular}{@{}cccc}
\br
$\rho$            			       & Steps (t)& Site (x) & Coin state \\ \mr
\multirow{2}{*}{\ \ \ $\frac{1}{4} \ \bigg\{ $} & $6$      & 2        & $-\beta\ket{\ua}+\alpha\ket{\da}$ \\
                               & $12$     & 1        & $\ket{\psi_{0,1}}$ \\
\multirow{2}{*}{\ \  \ $\frac{1}{2} \ \bigg\{ $} & $4$      & 2        & $-\beta\ket{\ua}+\alpha\ket{\da}$ \\ 
                               & $8$      & 1        & $\ket{\psi_{0,1}}$ \\ 
			$\frac{3}{4}$		         & $6$      & 1        & $\ket{\psi_{0,1}}$ \\
\br
\end{tabular}
\end{indented}
\end{table}

\noindent
After t-steps, we apply appropriate recovery operator, $(\rho',\theta',\phi')=(0,0,-\pi)$, on (\ref{eq:wf2line}). In this way, we obtain the initial coin state and hence PST. Overall process can be written as

\[
\ket{\psi_{0,1}}\otimes\ket{1} \ \ora{\mathbf{C}_R\mathbf{U}^t} \ \ket{\psi_{0,1}}\otimes\ket{2}.
\]

\noindent
Thus, any coin state can be transferred on 2-line perfectly with appropriate $(\rho, t)$ values given in table \ref{tb:2line}. We note that recovery operator is constant for a given coin operator and it provides PST for all initial coin states. In each PST case, the QW is periodic. For example, after 4 steps of the walk with $\rho=1/2$, if the walker proceeds $4$ more steps, the initial quantum state is recovered. There is also a case with $\rho=3/4$, where 2-line is periodic but it does not lead to PST. PST requires the total state to localize more than one sites in turn and this process naturally gives rise to periodicity.

After applying the recovery operator, we initialize the walker with the initial coin state at a different site. For example, when $\rho = 1/2$, if the walker is acted on by sequence of operations, such as $\mathbf{C}_R\mathbf{U}^4\mathbf{C}_R\mathbf{U}^4$, it will be initialized on sites $1$ and $2$ alternatingly. The sequence of initializations which keeps the initial coin state unchanged, suggest us to define a new classification for discrete-time QWs which we call \textit{n-periodicity}. We can define one step of the walk for the example above as $\textbf{U}'=\textbf{C}_R\textbf{U}^4$. Then, after each step, coin state will be conserved and the only change will occur in the position space. In other words, $\textbf{U}'$ is same as that of $\textbf{I}\otimes(\ket{2}\bra{1}+\ket{1}\bra{2})$. Since the walker is localized on two sites in an alternating manner, the QW under consideration becomes 2-periodic. In general, the number $n$ gives the total number of sites where initial coin state is localized during the time evolution. If QW is periodic, we will call it 1-periodic, i.e., well-known periodicity concept becomes a member of the general n-periodicity class. Thus, N-line or N-cycle allow maximum N-periodicity. This definition is useful because it generalizes the periodicity definition so that it includes the PST too.

For $\rho\neq1$, reflecting boundaries ensure that there will always be a non-zero probability for finding the walker at $A$, independent of $t$, if there is no destructive interference. However, the dimension of the position space for 2-line allows the wave function to vanish at $A$ and gives rise to the cases given in table \ref{tb:2line}. 

\subsection{\label{sec:nline2}Case: \texorpdfstring{$\rho=1$}{rho=1}}
When we restrict the coin operators to $\rho=1$, independent of the initial coin state, the walker is transferred from $A$ to $B$ and $B$ to $A$ at intervals of $N$ steps for all N-lines. In general, the walker is at the position $B$ or $A$ if $t=N(2l-1)$ or  $t=2Nl$ steps are taken, respectively. Here, $l \in \mathcal{Z}^+$ specifies the number of "round trips" of the walker within the lattice. To find the coin state of the walker at $t$, we have derived the total quantum states

\begin{equation}
\eqalign{
\ket{\Psi_{N(2l-1)}}& = e^{i(l-1)\Theta}[-\beta e^{i(\theta+\phi)}\ket{\ua}+\alpha \ket{\da}]\otimes\ket{N}, \cr
\ket{\Psi_{2Nl}}& = e^{il\Theta}[\alpha\ket{\ua}+\beta \ket{\da}]\otimes\ket{1},
}
\label{eq:nline1} 
\end{equation}

\noindent
where $\Theta(\theta,\phi,N)=(\theta+\phi) N + \mu\pi$ and $\theta,\phi$ are the parameters of the coin operator. Here, $\mu$ is a function which adds the phase $\pi$ for odd $N$ and it can be defined as $\mu(N)=[1-(-1)^N]/2$. It is shown in  (\ref{eq:nline1}) that the total state is periodically localized at opposite sites which agrees with the numerical results. Furthermore,  the walk is periodic with a period of $2Nl$ steps up to an overall phase. After $N$ steps, we apply recovery operator $(\rho',\theta',\phi')=(0,0,-\theta-\phi-\pi)$ for PST. Recovery operator is a function of $\theta$ and $\phi$ which means that for all coin operators with $\rho=1$, there is always a corresponding recovery operator. Hence, step operator $\mathbf{U}'=\mathbf{C}_R\mathbf{U}^N$ makes N-line 2-periodic for $\rho=1$.

\section{\label{sec:ncycle}PST on N-cycles}
\subsection{\label{sec:ncycle1}Case: \texorpdfstring{$\rho\neq1$}{rho!=1}}

For 2-cycle, full evolution can simply be written in matrix form as

\begin{equation}
\mathbf{U}^t \leftrightarrow
\left(
	\begin{array}{cc}
	\sqrt{\rho}				      &	 \sqrt{1-\rho}e^{i\theta}		   \\
	\sqrt{1-\rho}e^{i\phi}	&	-\sqrt{\rho}e^{i(\theta+\phi)} \\
	\end{array} 
\right)^{t}\otimes 
\left(
	\begin{array}{cc}
	0 &	1	\\
	1	&	0 \\
	\end{array}
\right)^t.
\label{eq:2cycleevol}
\end{equation}

\noindent
In (\ref{eq:2cycleevol}), we see that, shift operator swaps the position of the walker independent of its coin state. At $t=1$, the total state becomes (\ref{eq:coineffect}) with $x=2$. Since the coin operator is unitary, $\mathbf{C}_R=\mathbf{C}^{\dagger}\otimes\mathbf{I}$ leads to PST after first step. If we define one-step as $\mathbf{U}'=(\mathbf{C}^{\dagger}\otimes\mathbf{I})\mathbf{S}(\mathbf{C}\otimes\mathbf{I})$, QW becomes 2-periodic and it keeps the initial coin state unchanged. In other words, the initial coin state bounces back and forth between two sites. In contrast to 2-line, 2-cycle allows PST for all coin operators with the aid of appropriate recovery operators. We note that, if we choose $\theta=\phi=0$, without any recovery operator, the walk is naturally periodic with a period of $2$ steps for $\rho \in [0,1]$ which generalizes the $\rho=1/2$ condition in \cite{41}.

\begin{figure}[ht]
\centering
\includegraphics[scale=1.2]{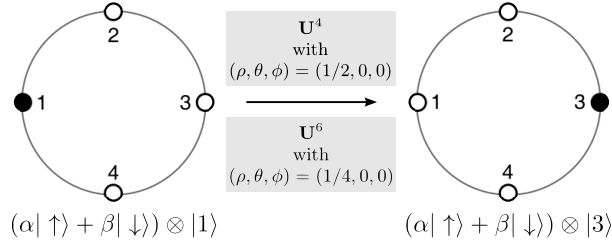}
\caption{PST on 4-cycle. This is the only case where discrete-time QW allows PST with Hadamard coin operator or with a biased coin, $\rho=1/4$, without any recovery operators. Black (hollow) dots indicate $P_{t,x}=1$ ($P_{t,x}=0$).}
\label{fig:4cycle}
\end{figure}

A special case for PST on circles is the 4-cycle. In this case, we achieve PST by using well-known Hadamard coin operator (in $4$ steps) or the biased coin operator $\rho=1/4$ (in $6$ steps) without any recovery operators. These results are given in  table \ref{tb:4cycle} and figure \ref{fig:4cycle}. We see that for each PST case the walk is also periodic. In \cite{41}, it has been already shown that 4-cycle has a period of $8$ steps for $\rho=1/2$. We extend this result by showing that it also has a period of $12$ steps for $\rho=1/4$ and period of $6$ steps for $\rho=3/4$.

\begin{table}[h]
\caption{\label{tb:4cycle}For the 4-cycle, these are the cases where the walker is found with probability 1. The other parameters of coin operator are chosen as $\theta=\phi=0$. The overall phase $e^{i\pi}$ for $\rho=1/2$ appears if $\theta=\pi$.}
\begin{indented}
\item[]\begin{tabular}{@{}cccc}
\br
$\rho$            			   & Step(t)  &  Site(x) & Coin state \\ \mr
\multirow{2}{*}{\ \ \ $\frac{1}{4} \ \bigg\{ $} & $6$      & 3 & $\ket{\psi_{0,1}}$ \\ 
                               & $12$     & 1 & $\ket{\psi_{0,1}}$ \\ 
\multirow{2}{*}{\ \ \ $\frac{1}{2} \ \bigg\{ $} & $4$      & 3 & $(e^{i\pi})\ket{\psi_{0,1}}$ \\ 
                               & $8$      & 1 & $\ket{\psi_{0,1}}$ \\ 
$\frac{3}{4}$&6&1&$\ket{\psi_{0,1}}$ \\
\br
\end{tabular}
\end{indented}
\end{table}

\subsection{\label{sec:ncycle2}Case: \texorpdfstring{$\rho=1$}{rho=1}} 
Now, we consider the N-cycles with even $N$ and $\theta,\phi\neq0$. Since the coin operator is diagonal, $\ket{\ua}$ and $\ket{\da}$ terms do not mix, and generate propagations in opposite directions. After $N/2$ steps, we find the walker at $B$ with probability $1$. We note that the coin operator adds the phase $e^{i(\theta+\phi+\pi)}$ to the coefficient of $\ket{\da}$  in each step. Thus, after $N/2$ steps, the total state becomes

\[
\ket{\mathbf{\Psi}_{N/2}}=(\alpha\ket{\ua}+\beta e^{i\frac{N\Theta}{2}}\ket{\da})\otimes\ket{\frac{N}{2}+1},
\]

\noindent
where $\Theta=\theta+\phi+\pi$. Without loss of generality, one can choose $\theta'=0$ and use the recovery operator $(\rho',\theta',\phi')=(1,0,-[N\Theta/2]+\pi)$ to achieve PST. The step operator $\mathbf{U}' = \mathbf{C}_R\mathbf{U}^{N/2}$ makes the walk 2-periodic. If $N$ is odd, wave function does not localize at any site except the initial one. The total state after $N$ steps is

\[
\ket{\mathbf{\Psi}_{N}}=(\alpha\ket{\ua}+\beta e^{iN\Theta}\ket{\da})\otimes\ket{1}.
\]

\noindent
Appropriate choice for the recovery operator can be given as $(\rho',\theta',\phi')=(1,0,-N\Theta+\pi)$ at $t=N$. The step operator $\mathbf{U}'=\mathbf{C}_R\mathbf{U}^N$ makes the walk 1-periodic.

We have shown that the coin operator which is restricted to $\rho=1$ allows PST on N-cycles. Although, it has not been indicated in the discussion about N-cycles above, spatial approach has been used intrinsically, i.e., clockwise rotations correspond to $\ket{\ua}$. If the walk is driven by the coin operator $(\rho,\theta,\phi)=(0,0,0)$ (the flip coin operator), we define the directions with the local approach for PST (see figure \ref{fig:dirappr}). When $N$ is odd, N-cycle is ill-defined since we have to label at least two edges with the same basis state. Therefore, we consider N-cycles with even $N$ only. If we label all edges as in Fig. \ref{fig:dirappr}(b), after $N/2$ steps, the total state becomes

\[ 
\ket{\mathbf{\Psi}_{N/2}}=
\cases{\ket{\psi_{0,in}}\otimes\ket{\frac{N}{2}+1}&for even $ N/2$\\
(\alpha e^{i\phi}\ket{\da}+\beta e^{i\theta}\ket{\ua})\otimes\ket{\frac{N}{2}+1}&for odd $N/2$\\}.
\]

\noindent
Both case have the overall phase $e^{i \left \lfloor{N/4}\right \rfloor (\theta+\phi)}$   where $\left \lfloor{}\right \rfloor$ is the floor function. The first case shows that PST is achieved after $N/2$ steps. Also, it is clear that we can use $(\rho',\theta',\phi')=(0,-\phi,-\theta)$ to recover the second case and make the walk periodic. However, N-lines do not have the same property, i.e., PST is not possible if we use flip coin operator with local approach. We can demonstrate this fact by evaluating the first two steps as follows:

\[
\ket{\Psi_0}=\alpha\ket{\ua 1}+\beta\ket{\da 1}\ora{\mathbf{U}}\alpha\ket{\ua 1}+\beta\ket{\da 2}\ora{\mathbf{U}}\alpha\ket{\ua 1}+\beta\ket{\da 3} \ora{\mathbf{U}} \cdots
\]

\noindent
where $\mathbf{U}$ involves the flip coin operator. We see that after each step, the first term in the summation is stuck at site $1$ because of the periodicity condition. For an N-line, after $N$ steps, the second term will be stuck at site $N$ as well. Thus, neither PST nor periodicity is possible.

\section{\label{sec:conc}CONCLUSION}
We have proposed a PST scheme by introducing recovery operators in discrete-time QW architecture on N-lines and N-cycles. We have shown that by using identity or flip coin operator, an unknown qubit state can be transferred to an arbitrary distance perfectly with the aid of appropriate recovery operator. The 2-cycle is the only lattice which allows PST for all coin operators up to N=10. Also, the Hadamard coin and biased coin $\rho=1/4$ allow PST on 2-line. We have shown that the 4-cycle is a special case where PST occurs if the walk is driven by the Hadamard coin operator or the biased coin operator $\rho=1/4$, without any recovery operators. Moreover, we have introduced new periodic discrete-time QWs on N-lines and also extended periodicity cases which has already been known for 2-cycle and 4-cycle \cite{41}. We have shown the strong relation between the periodicity and PST.

Since recovery operators are just additional coin operators and PST occurs after small number of steps (which is comparable with the lattice size), it seems that the experimental realization of our scheme is quite feasible with today's technology.

\section*{ACKNOWLEDGEMENTS}
We would like to thank B. Pekerten, G. Karpat and B. \c{C}akmak for helpful discussions. This work has been partially supported by the Scientific and Technological Research Council of Turkey (TUBITAK) under Grant No. 111T232.


\section*{References}
\begin{thebibliography}{42}

\bibitem{1}
Nikolopoulos G M and Jex I 2014 \textit{Quantum State Transfer and Network Engineering} (Berlin: Springer)

\bibitem{2} 
Bose S 2003 \PRL \textbf{91} 207901

\bibitem{3}
Subrahmanyam V 2004 \PR A \textbf{69} 034304

\bibitem{4}
W\'{o}jcik A, \L uczak T, Kurzy\'{n}ski P, Grudka A, Gdala T and Bednarska M 2007 \PR A \textbf{75} 022330

\bibitem{5}
Di Franco C, Paternostro M, and Kim M S 2008 \PRL \textbf{101} 230502

\bibitem{6}
Chudzicki C and Strauch F W 2010 \PRL \textbf{105} 260501

\bibitem{7}
Lemr K, Bartkiewicz K, Cernoch A and Soubusta J 2013 \PR A \textbf{87} 062333

\bibitem{8}
Paganelli S, Lorenzo S, Apollaro T J G, Plastina F and Giorgi G L 2013 \PR A \textbf{87}, 062309

\bibitem{9}
Zwick A, \'{A}lvarez G A, Stolze J and Osenda O 2011 \PR A \textbf{84}, 022311

\bibitem{10}
Zwick A, \'{A}lvarez G A, Stolze J and Osenda O 2012 \PR A \textbf{85}, 012318

\bibitem{11}
Christandl M, Datta N, Ekert A and Landahl A J 2004 \PRL \textbf{92}, 187902

\bibitem{12}
Christandl M, Datta N, Dorlas T C, Ekert A, Kay A and Landahl A J 2005 \PR A \textbf{71}, 032312

\bibitem{13}
Farhi E and Gutmann S 1998 \PR A \textbf{58} 915

\bibitem{14}
Aharonov Y, Davidovich L and Zagury N 1993 \PR A \textbf{48} 1687

\bibitem{15}
Kempe J 2003 \textit{Contemp. Phys.} \textbf{44} 307

\bibitem{16}
Travaglione B C and Milburn G J 2002 \PR A \textbf{65} 032310

\bibitem{17}
D\"{u}r W, Raussendorf R, Kendon V M and Briegel H J 2002 \PR A \textbf{66} 052319

\bibitem{18}
Sanders B C, Bartlett S D, Tregenna B and Knight P L 2003 \PR A \textbf{67} 042305

\bibitem{19}
Xue P, Sanders B C and Leibfried D 2009 \PRL \textbf{103} 183602

\bibitem{20}
Bouwmeester D, Marzoli I, Karman G P, Schleich W and Woerdman J P 1999 \PR A \textbf{61} 013410

\bibitem{21}
Du J, Li H, Xu X, Shi M, Wu J, Zhou X and Han R 2003 \PR A \textbf{67} 042316

\bibitem{22}
C\^{o}t\'{e} R, Russell A, Eyler E E and Gould P L 2006 \NJP \textbf{8} 156

\bibitem{23}
Perets H B, Lahini Y, Pozzi F, Sorel M, Morandotti R and Silberberg Y 2008 \PRL \textbf{100} 170506

\bibitem{24}
Karski M, F\"{o}rster L, Choi J M, Steffen A, Alt W, Meschede D and Widera A 2009 \textit{Science} \textbf{325} 174

\bibitem{25}
Schmitz H, Matjeschk R, Schneider Ch, Glueckert J, Enderlein M, Huber T and Schaetz T 2009 \PRL \textbf{103} 090504

\bibitem{26}
Schreiber A, Cassemiro K N, Poto\v{c}ek V, G\'{a}bris A, Mosley P J, Andersson E, Jex I and Silberhorn Ch 2010 \PRL \textbf{104} 050502

\bibitem{27}
Z\"{a}hringer F, Kirchmair G, Gerritsma R, Solano E, Blatt R and Roos C F 2010 \PRL \textbf{104} 100503

\bibitem{28}
Broome M A, Fedrizzi A, Lanyon B P, Kassal I, Aspuru-Guzik A and White A G 2010 \PRL \textbf{104} 153602

\bibitem{29}
Peruzzo A, Lobino M, Matthews J C F, Matsuda N, Politi A, Poulios K, Zhou X Q, Lahini Y, Ismail N, W\"{o}rhoff K, Bromberg Y, Silberberg Y, Thompson M G and O’Brien J L 2010 \textit{Science} \textbf{329}, 1500

\bibitem{30}
Sansoni L,  Sciarrino F, Vallone G, Mataloni P, Crespi A, Ramponi R and Osellame R 2012 \PRL \textbf{108} 010502

\bibitem{31}
Schreiber A, G\'{a}bris A, Rohde P P, Laiho K, \v{S}tefa\v{n}\'{a}k M, Poto\v{c}ek V, Hamilton C, Jex I and Silberhorn C 2012 \textit{Science} \textbf{336} 55

\bibitem{32}
Ghosh J 2014 \PR A \textbf{89} 022309

\bibitem{33}
Kurzy\'{n}ski P and W\'{o}jcik A 2011 \PR A \textbf{83} 062315

\bibitem{34}
Kendon V M and Tamon C 2011 \textit{J. Comp. Theor. Nanoscience} \textbf{8} 422

\bibitem{35}
Barr K, Proctor T, Allen D and Kendon V 2014 \textit{Quant. Inf. and Comp.} \textbf{14} 417-38

\bibitem{36}
Romanelli A, Siri R, Abal G, Auyuanet A and Donangelo R 2005 \textit{Physica} A \textbf{347} 137–52

\bibitem{37}
P\'{o}lya G 1921 \textit{Mathematische Annalen} \textbf{84} 149

\bibitem{38}
\v{S}tefa\v{n}\'{a}k M, Jex I and Kiss T 2008 \PRL \textbf{100} 020501

\bibitem{39}
\v{S}tefa\v{n}\'{a}k M, Kiss T and Jex I 2009 \NJP \textbf{11} 043027
 
\bibitem{40}
Rohde P P, Fedrizzi A and Ralph T C 2012 \textit{J. Mod. Opt.} \textbf{59} 8 710-20

\bibitem{41}
Tregenna B, Flanagan W, Maile R and Kendon V 2003 \NJP \textbf{5} 83.1

\bibitem{42}
Dukes P R 2014 \textit{Results Phys.} \textbf{4} 189-97

\bibitem{43}
Zhan X, Qin H, Bian Z H, Li J and Xue P 2014 \PR A \textbf{90} 012331
\end{thebibliography}
\end{document}